\documentclass[12pt]{article}
\usepackage{amssymb,amscd,array}
\catcode `\@=11
\@addtoreset{equation}{section}

\newtheorem{thm}{Theorem}[section]
\newtheorem{rem}[thm]{Remark}

\def\qed{\blacksquare}
\newcommand{\be}{\begin{equation}}
\newcommand{\ee}{\end{equation}}
\newcommand{\bea}{\begin{eqnarray}}
\newcommand{\eea}{\end{eqnarray}}
\newcommand{\R}{\mathbb{R}}
\newcommand{\N}{\mathbb{N}}
\newcommand{\C}{\mathbb{C}}
\newcommand{\p}{\partial}

\begin{document}
\begin{titlepage}

\begin{center}
{\bf \Large{On the Super-Renormalizablity of Quantum Gravity in the Linear Approximation\\}}
\end{center}
\vskip 1.0truecm
\centerline{D. R. Grigore, 
\footnote{e-mail: grigore@theory.nipne.ro}}
\vskip5mm
\centerline{Department of Theoretical Physics}
\centerline{Institute for Physics and Nuclear Engineering ``Horia Hulubei"}
\centerline{Bucharest-M\u agurele, P. O. Box MG 6, ROM\^ANIA}

\vskip 2cm
\bigskip \nopagebreak
\begin{abstract}
\noindent
We compute the one-loop contributions of the chronological products for 
massless gravity in the second order of the perturbation theory. We prove
that the loop contributions are coboundaries i.e. expressions which give zero
when averaged on physical states. We conjecture that such a result should be 
true in higher orders of the perturbation theory also. 
This result should make easier the problem of constructive quantum field
theory.
\end{abstract}
\end{titlepage}

\section{Introduction}

We remind a few facts about perturbative quantum field theory following essentially \cite{cohomology}.
The general framework of perturbation theory consists in the construction of 
the chronological products: for every set of Wick monomials 
$ 
W_{1}(x_{1}),\dots,W_{n}(x_{n}) 
$
acting in some Fock space
$
{\cal H}
$
one associates the operator
$ 
T^{W_{1},\dots,W_{n}}(x_{1},\dots,x_{n}); 
$  
all these expressions are in fact distribution-valued operators called
chronological products. It will be convenient to use another notation: 
$ 
T(W_{1}(x_{1}),\dots,W_{n}(x_{n})). 
$ 
These operators are constrained by Bogoliubov axioms  \cite{BS}, \cite{EG},
\cite{DF} and their construction can be done recursively according to Epstein-Glaser
prescription \cite{EG} (which reduces the induction procedure to a
distribution splitting of some distributions with causal support). These products are not uniquely
defined but there are some natural limitation on the arbitrariness. This limitation
is a bound on the degree of the singularity of the vacuum averages of the chronological
products. One imposes that this singularity degree should be as low as possible.
If this arbitrariness does not grow with the order $n$ of the perturbation theory then
we say that the theory is renormalizable; the most popular point of view is that
only such theories are physically meaningful. Apparently, this power counting argument
excludes quantum gravity. We will argue that there is a way out of this no-go result.
The basic idea is that quantum gravity must be considered, at least in the perturbation
theory, as a theory of particles of zero mass and helicity $2$. Such theories are best
described using physical and non-physical fields (the ghost fields) as well. This means that the Fock space will
contain physical and non-physical states. Therefore one should investigate the 
ultraviolet behavior of the chronological products {\it resticted} to the subspace 
of physical states. If we do that we can prove by direct computation that the one-loop 
contributions are in fact null if we restrict ourselves to the subspace of physical states.
We have given in \cite{sr2} a pure cohomological argument of this assertion. Here we use 
a new approach based on explicit computations of the loop contributions and explicit
proof of their triviality. 

From the technical point of view, one must first construct a Fock space
$
{\cal H}
$
with indefinite metric, generated by physical and un-physical fields (called
{\it ghost fields}). One selects the physical states assuming the existence of
an operator $Q$ called {\it gauge charge} which verifies
$
Q^{2} = 0
$
and such that the {\it physical Hilbert space} is by definition
$
{\cal H}_{\rm phys} \equiv Ker(Q)/Im(Q).
$

One has a natural grading in the Hilbert space  
$
{\cal H}
$
and the graded commutator
$
d_{Q}
$
of the gauge charge with any operator $A$ is defined by
\be
d_{Q}A = [Q,A]
\ee
where in the right hand side we understand that
$
[\cdot,\cdot]
$
is the graded commutator. Because
\be
Q^{2} = 0 \qquad \Longleftrightarrow \qquad d_{Q}^{2} = 0
\ee
it means that
$
d_{Q}
$
is a co-chain operator.
 
A gauge theory assumes also that there exists a Wick polynomial of null ghost
number
$
T(x)
$
called {\it the interaction Lagrangian} such that
\be
~[Q, T] = i \partial_{\mu}T^{\mu}
\label{gau1}
\ee
for some other Wick polynomials
$
T^{\mu}.
$
This relation means that the expression $T$ leaves invariant the physical
states, at least in the adiabatic limit. Indeed, if this is true we have:
\be
T(f)~{\cal H}_{\rm phys}~\subset~~{\cal H}_{\rm phys}  
\label{gau2}
\ee
up to terms which can be made as small as desired (making the test function $f$
flatter and flatter). In the case of quantum gravity we also have the Wick polynomials
$
T^{\mu},~T^{\mu\nu},~T^{\mu\nu\rho}
$
such that:
\be
~[Q, T] = i \partial_{\mu}T^{\mu}, \quad
[Q, T^{\mu}] = i \partial_{\nu}T^{\mu\nu}, \quad
[Q, T^{\mu\nu}] = i \partial_{\rho}T^{\mu\nu\rho} \quad 
[Q, T^{\mu\nu\rho}] = 0 
\label{descent}
\ee
and the expressions
$
T^{\mu\nu},~T^{\mu\nu\rho}
$
are completely antisymmetric in all indexes so we can also use
a compact notation
$
T^{I}
$
where $I$ is a collection of indexes
$
I = [\nu_{1},\dots,\nu_{p}]~(p = 0,1,\dots,)
$
and the brackets emphasize the complete antisymmetry in these indexes. All these
polynomials have the same canonical dimension
\be
\omega(T^{I}) = \omega_{0} = 5,~\forall I
\ee
and because the ghost number of
$
T \equiv T^{\emptyset}
$
is supposed null, then we also have:
\be
gh(T^{I}) = |I|.
\ee
One can write compactly the relations (\ref{descent}) as follows:
\be
d_{Q}T^{I} = i~\partial_{\mu}T^{I\mu}.
\label{descent1}
\ee
If the interaction Lagrangian $T$ is Lorentz invariant, then one can prove that
the expressions
$
T^{I},~|I| > 0
$
can be taken Lorentz covariant as well.

Now we can construct the chronological products
$$
T^{I_{1},\dots,I_{n}}(x_{1},\dots,x_{n}) \equiv
T(T^{I_{1}}(x_{1}),\dots,T^{I_{n}}(x_{n}))
$$
according to the recursive procedure of Epstein and Glaser. We say that the theory is gauge invariant
in all orders of the perturbation theory if the following set of identities
generalizing (\ref{descent1}):
\be
d_{Q}T^{I_{1},\dots,I_{n}} = 
i \sum_{l=1}^{n} (-1)^{s_{l}} {\partial\over \partial x^{\mu}_{l}}
T^{I_{1},\dots,I_{l}\mu,\dots,I_{n}}
\label{gauge}
\ee
are true for all 
$n \in \N$
and all
$
I_{1}, \dots, I_{n}.
$
Here we have defined
\be
s_{l} \equiv \sum_{j=1}^{l-1} |I|_{j}.
\ee

We introduce some cohomology terminology. We consider a {\it cochains} to be
an ensemble of distribution-valued operators of the form
$
C^{I_{1},\dots,I_{n}}(x_{1},\dots,x_{n}),~n = 1,2,\dots
$
(usually we impose some supplementary symmetry properties) and define the
derivative operator $\delta$ according to
\be
(\delta C)^{I_{1},\dots,I_{n}}
= \sum_{l=1}^{n} (-1)^{s_{l}} {\partial\over \partial x^{\mu}_{l}}
C^{I_{1},\dots,I_{l}\mu,\dots,I_{n}}.
\ee
We can prove that 
\be
\delta^{2} = 0.
\ee
Next we define
\be
s = d_{Q} - i \delta,\qquad \bar{s} = d_{Q} + i \delta
\ee
and note that
\be
s \bar{s} = \bar{s} s = 0.
\ee
We call {\it relative cocycles} the expressions $C$ verifying
\be
sC = 0
\ee
and a {\it relative coboundary} an expression $C$ of the form
\be
C = \bar{s}B.
\ee
The relation (\ref{gauge}) is simply the cocycle condition
\be
sT = 0.
\ee

The purpose of this paper is to investigate if this condition implies that, at
least some contributions of $T$, are in fact coboundaries. Coboundaries are 
trivial from the physical point of view: if we consider two physical states
$
\Psi, \Psi^{\prime}
$
then
\bea
<\Psi, \bar{s}B\Psi^{\prime}> = 
< Q\Psi, B\Psi^{\prime}> - < \Psi, BQ\Psi^{\prime}>
+ i < \Psi, \delta B\Psi^{\prime}> 
\nonumber\\
= i < \Psi, \delta B\Psi^{\prime}> \rightarrow 0
\eea
(in the adiabatic limit). 

We will consider here only the second order of the perturbation theory and prove
that for  massless gravity the loop contributions is a coboundary i.e. the theory is
essentially classical. This follows from the fact that in the loop expansion
the $0$-loop (or tree) contribution corresponds to the classical theory
\cite{DF3}.

In the next Section we present the description of the free fields we use, mainly to
fix the notations. In Section \ref{ggt} we remind Bogoliubov axioms for the second
order of the perturbation theory and we give the basic
distributions with causal support appearing for loop contributions in the second
order of the perturbation theory. In Section \ref{gravity} we
prove the cohomology result for gravity.
\newpage

\section{Massless Particles of Spin $2$ (Gravitons)}

We refer to more details to \cite{cohomology}. We consider the vector space 
$
{\cal H}
$
of Fock type generated (in the sense of Borchers theorem) by the symmetric
tensor field 
$
h_{\mu\nu}
$ 
(with Bose statistics) and the vector fields 
$
u^{\rho}, \tilde{u}^{\sigma}
$
(with Fermi statistics). We suppose that all these (quantum) fields are of null mass. 
In this vector space we can define a sesquilinear form 
$<\cdot,\cdot>$
in the following way: the (non-zero) $2$-point functions are by definition:
\bea
<\Omega, h_{\mu\nu}(x_{1}) h_{\rho\sigma}(x_{2})\Omega> = - {i\over 2}~
(\eta_{\mu\rho}~\eta_{\nu\sigma} + \eta_{\nu\rho}~\eta_{\mu\sigma}
- \eta_{\mu\nu}~\eta_{\rho\sigma})~D_{0}^{(+)}(x_{1} - x_{2}),
\nonumber \\
<\Omega, u_{\mu}(x_{1}) \tilde{u}_{\nu}(x_{2})\Omega> = i~\eta_{\mu\nu}~
D_{0}^{(+)}(x_{1} - x_{2}),
\nonumber \\
<\Omega, \tilde{u}_{\mu}(x_{1}) u_{\nu}(x_{2})\Omega> = - i~\eta_{\mu\nu}~
D_{0}^{(+)}(x_{1} - x_{2})
\eea
and the $n$-point functions are generated according to Wick theorem, or equivalently assuming that the truncated
$n$-point functions are null for 
$
n \geq 3.
$
Here
$
\eta_{\mu\nu}
$
is the Minkowski metrics (with diagonal $1, -1, -1, -1$) and 
$
D_{0}^{(+)}
$
is the positive frequency part of the Pauli-Jordan distribution
$
D_{0}
$
of null mass. To extend the sesquilinear form to
$
{\cal H}
$
we define the conjugation by
\be
h_{\mu\nu}^{\dagger} = h_{\mu\nu}, \qquad 
u_{\rho}^{\dagger} = u_{\rho}, \qquad
\tilde{u}_{\sigma}^{\dagger} = - \tilde{u}_{\sigma}.
\ee

Now we can define in 
$
{\cal H}
$
the operator $Q$ according to the following formulas:
\bea
~[Q, h_{\mu\nu}] = - {i\over 2}~(\partial_{\mu}u_{\nu} + \partial_{\nu}u_{\mu}
- \eta_{\mu\nu} \partial_{\rho}u^{\rho}),\qquad
[Q, u_{\mu}] = 0,\qquad
[Q, \tilde{u}_{\mu}] = i~\partial^{\nu}h_{\mu\nu}
\nonumber \\
Q\Omega = 0
\label{Q-0-2}
\eea
where by 
$
[\cdot,\cdot]
$
we mean the graded commutator. One can prove that $Q$ is well defined. Indeed,
we have the causal commutation relations 
\bea
~[h_{\mu\nu}(x_{1}), h_{\rho\sigma}(x_{2}) ] = - {i\over 2}~
(\eta_{\mu\rho}~\eta_{\nu\sigma} + \eta_{\nu\rho}~\eta_{\mu\sigma}
- \eta_{\mu\nu}~\eta_{\rho\sigma})~D_{0}(x_{1} - x_{2})~\cdot I,
\nonumber \\
~[u(x_{1}), \tilde{u}(x_{2})] = i~\eta_{\mu\nu}~D_{0}(x_{1} - x_{2})~\cdot I
\eea
and the other commutators are null. The operator $Q$ should leave invariant
these relations, in particular 
\be
[Q, [ h_{\mu\nu}(x_{1}),\tilde{u}_{\sigma}(x_{2})]] + {\rm cyclic~permutations}
= 0
\ee
which is true according to the preceding relations. Then we have:
\begin{thm}
The operator $Q$ verifies
$
Q^{2} = 0.
$ 
The factor space
$
Ker(Q)/Im(Q)
$
is isomorphic to the Fock space of particles of zero mass and helicity $2$
(gravitons). 
\label{fock-0}
\end{thm}
If we define
\be
h \equiv \eta_{\mu\nu}~ h^{\mu\nu}
\ee
we also have
\bea
~[h_{\mu\nu}(x_{1}), h(x_{2}) ] = [h(x_{1}), h_{\mu\nu}(x_{2}) ]= i~\eta_{\mu\nu}~\cdot I,
\nonumber \\
~[h(x_{1}), h(x_{2}) ] = 4 i~D_{0}(x_{1} - x_{2})~\cdot I
\eea
and
\be
d_{Q} h = i~\partial_{\mu}u^{\mu}.
\ee
\newpage
\section{General Gauge Theories\label{ggt}}
 
We give here the essential ingredients of perturbation theory for the order
$
n = 2
$
of the perturbation theory. We asignate to
$ 
h_{\mu\nu}, u_{\mu}, \tilde{u}_{\mu}
$
the canonical dimension
$
1
$.
A derivative applied to a field raises the canonical dimension by $1$. The ghost
number of 
$ 
h_{\mu\nu}
$
is $0$ and for the ghost fields 
$ 
u_{\mu}, \tilde{u}_{\mu}
$
is $1$. The Fermi parity of a Fermi (Bose) field is $1$ (resp. $0$). The canonical 
dimension of a Wick monomial is additive with respect to the factors and the
same is true for the ghost number and the Fermi parity.
\subsection{Bogoliubov Axioms}{\label{bogoliubov}}

Suppose that the Wick monomials
$
A, B
$
are self-adjoint:
$
A^{\dagger} = A,~B^{\dagger} = B
$
and of fixed Fermi parity 
$
|A|, |B|
$
and canonical dimension
$
\omega(A), \omega(B)
$.
For gravity we must take
$
\omega(A), \omega(B) = 5
$.
The chronological products
$ 
T(A(x),B(y))
$
are verifying the following set of axioms:
\begin{itemize}
\item
Skew-symmetry:
\be
T(B(y),A(x)) = (-1)^{|A||B| } T(A(x),B(y))
\ee
\item
Poincar\'e invariance: we have a natural action of the Poincar\'e group in the
space of Wick monomials and we impose that for all elements $g$ of the
universal 
covering group
$
inSL(2,\C)
$
of the Poincar\'e group:
\be
U_{g} T(A(x),B(y)) U^{-1}_{g} = T(g\cdot A(g\cdot x),g\cdot B(g\cdot y))
\label{invariance}
\ee
where
$
x \mapsto g\cdot x
$
is the action of
$
inSL(2,\C)
$
on the Minkowski space. 
\item
Causality: if
$
x \succeq y
$
i.e.
$
y \cap ( x + \bar{V}^{+} ) = \emptyset
$
then we have:
\be
T(A(x),B(y)) = A(x)~B(y);
\label{causality}
\ee
\item
Unitarity: If we define the {\it anti-chronological products} according to
\be
\bar{T}(A(x),B(y)) \equiv A(x) B(y) + B(y) A(x) - T(A(x),B(y)) 
\label{antichrono}
\ee
then the unitarity axiom is:
\be
\bar{T}(A(x),B(y) = T(A(x),B(y))^{\dagger}.
\label{unitarity}
\ee
\end{itemize}

It can be proved that this system of axioms can be supplemented with
\be
T(A(x),B(y)) = \sum \quad
<\Omega, T(A_{1}(x),B_{1}(y))\Omega>~:A_{2}(x)B_{2}(y):
\label{wick-chrono2}
\ee
where
$
A = A_{1}A_{2}, B = B_{1}B_{2}
$
is an arbitrary decomposition of $A$ and resp. $B$ in Wick submonomials and we
have supposed for simplicity that no Fermi fields are present; if Fermi fields 
are present, then some apropriate signs do appear. This is called the {\it Wick
expansion property}. 

We can also include in the induction hypothesis a limitation on the order of
singularity of the vacuum averages of the chronological products:
\be
\omega(<\Omega, T(A(x),B(y))\Omega>) \leq
\omega(A) + \omega(B) - 4
\label{power}
\ee
where by
$\omega(d)$
we mean the order of singularity of the (numerical) distribution $d$ and by
$\omega(W)$
we mean the canonical dimension of the Wick monomial $W$.

The contributions verifying
\be
\omega(<\Omega, T(A(x),B(y))\Omega>) <
\omega(A) + \omega(B) - 4
\label{superpower}
\ee
will be called {\it super-renormalizable}.

The operator-valued distributions
$
D, A, R, T
$
admit a decomposition into loop contributions
$
D = \sum _{l} D_{(l)}
$,
etc. Indeed every contribution is associated with a certain Feynman graph and
the 
integer $l$ counts the number of the loops. Alternatively, if we consider the
loop
decomposition of the advanced (or retarded) products we have in fact series in
$
\hslash
$
so the contribution corresponding to
$
l = 0
$
(the tree contribution) is the classical part and the loop contributions
$
l > 0
$
are the quantum corrections \cite{DF3}.
\newpage

\subsection{Second Order Chronological Products}
We go to the second order of perturbation theory using the {\it causal
commutator}
\be
D^{A,B}(x,y) \equiv D(A(x),B(y)) = [ A(x),B(y)]
\ee
where 
$
A(x), B(y)
$
are arbitrary Wick monomials and, as always we mean by 
$
[\cdot,\cdot]
$
the graded commutator. These type of distributions are translation invariant
i.e. they depend only on 
$
x - y
$
and the support is inside the light cones:
\be
supp(D) \subset V^{+} \cup V^{-}.
\ee

A theorem from distribution theory guarantees that one can causally split this
distribution:
\be
D(A(x),B(y)) = A(A(x),B(y)) - R(A(x),B(y)).
\ee
where:
\be
supp(A) \subset V^{+} \qquad supp(R) \subset V^{-}.
\ee
The expressions 
$
A(A(x),B(y)), R(A(x),B(y))
$
are called {\it advanced} resp. {\it retarded} products. They are not uniquely
defined: one can modify them with {\it quasi-local terms} i.e. terms
proportional with
$
\delta(x - y)
$
and derivatives of it. 

There are some limitations on these redefinitions coming from Lorentz
invariance, and {\it power counting}: this means that we should not make the
various distributions appearing in the advanced and retarded products too
singular.

Then we define the {\it chronological product} by:
\be
T(A(x),B(y)) = A(A(x),B(y)) + B(y) A(x) = R(A(x),B(y)) + A(x) B(y).
\ee
In the particular case of a gauge theory (as it is quantum gravity) we need that causal commutators
\be
D^{IJ}(x,y) \equiv [T^{I}(x), T^{J}(y)]
\ee
with the symmetry property
\be
D^{JI}(y,x) = - (-1)^{|I||J|}~D^{IJ}(x,y)
\ee
and the limitations
\be
gh(D^{IJ}) = |I|+ |J|
\ee
and power counting limitations coming from (\ref{power}). To get some explicit intuition, we need some details about (numerical)
distributions with causal support appearing in the second order of the perturbation theory.
\newpage
\subsection{Second Order Causal Distributions\label{causal}}

We remind the fact that the Pauli-Villars distribution is defined by
\be
D_{m}(x) = D_{m}^{(+)}(x) + D_{m}^{(-)}(x)
\ee
where 
\be
D_{m}^{(\pm)}(x) \sim 
\int dp e^{i p\cdot x} \theta(\pm p_{0}) \delta(p^{2} - m^{2})
\ee
such that
\be
D_{m}^{(-)}(x) = - D_{m}^{(+)}(- x).
\ee

This distribution has causal support. In fact, it can be causally split
(uniquely) into an
advanced and a retarded part:
\be
D_{m} = D_{m}^{\rm adv} - D_{m}^{\rm ret}
\ee
and then we can define the Feynman propagator and antipropagator
\be
D_{m}^{F} = D_{m}^{\rm ret} + D_{m}^{(+)}, \qquad \bar{D}_{m}^{F} = D_{m}^{(+)} - D_{m}^{\rm adv}.
\ee
All these distributions have singularity order
$
\omega(D) = -2
$.
We will consider from now on only the case
$
m = 0.
$

It is easy to see that the tree contribution in the second order of perturbation theory of the 
causal commutator is of the form 
\be
D_{(0)}(A(x),B(y)) = \sum p_{j}(\partial) D_{0}(x - y)~W_{j}(x,y)
\label{tree}
\ee
where 
$
p_{j}
$
are monomials in the partial derivatives and
$
W_{j}(x,y)
$
are some Wick monomials. Then we can obtain the advanced, retarded and chronological products by
simply replacing
$
D_{0}
$
by
$
D_{0}^{\rm adv}
$,
$
D_{0}^{\rm ret}
$
and
$
D_{0}^{F}
$
respectively. 

It is not hard to see that a formula similar to (\ref{tree}) is valid for the loop contributions
\be
D_{(1)}(A(x),B(y)) = \sum p_{j}(\partial) d_{2}(x - y)~W_{j}(x,y)
\label{one-loop}
\ee
where we will need the basic distribution
\be
d_{2}(x) \equiv {1\over 2} [ D_{0}^{(+)}(x)^{2} - D_{0}^{(+)}(- x)^{2} ]
\ee
which is also with causal support and it can be causally split as above in
\be
d_{2} = d_{2}^{\rm adv} - d_{2}^{\rm ret}
\label{d2}
\ee
and the corresponding Feynman propagators can be defined. These distributions
have the singularity order
$
\omega(D) = 0
$
so the causal splitting is not unique: we can add an arbitrary contribution of the form
$
\sim \delta(x).
$

In the explicit computations some associated distributions with causal
support do appear. We can have two derivatives distributed in two ways on the two factors
$
D_{0}^{(+)}
$:
\bea
d_{\mu\nu}(x) = D_{0}^{(+)}(x) \partial_{\mu}\partial_{\nu}D_{0}^{(+)}(x) 
- D_{0}^{(+)}(- x) \partial_{\mu}\partial_{\nu}D_{0}^{(+)}(-x)
\nonumber \\
f_{\mu\nu}(x) = \partial_{\mu}D_{0}^{(+)}(x) \partial_{\nu}D_{0}^{(+)}(x) 
- \partial_{\mu}D_{0}^{(+)}(- x) \partial_{\nu}D_{0}^{(+)}(-x).
\label{d2-1}
\eea
It is not hard to prove that we have
\be
d_{\mu\nu} = 
{2\over 3} \left( \partial_{\mu}\partial_{\nu} - {1\over 4} \square\right)d_{2}, \qquad
f_{\mu\nu} = 
{1\over 3} \left( \partial_{\mu}\partial_{\nu} + {1\over 2} \square\right)d_{2}.
\ee
We also have:
\be
f \equiv \eta^{\mu\nu}~f_{\mu\nu} = \square d_{2}
\ee

Now we introduce distributions with three derivatives:
\bea
d_{\{\mu\nu\rho\}} \equiv D^{(+)}_{0} \partial_{\mu}\partial_{\nu}\partial_{\rho}D^{(+)}_{0}
- D^{(-)}_{0} \partial_{\mu}\partial_{\nu}\partial_{\rho}D^{(-)}_{0}
\nonumber\\
f_{\{\mu\nu\}\rho} \equiv \partial_{\rho}D^{(+)}_{0} \partial_{\mu}\partial_{\nu}D^{(+)}_{0}
- \partial_{\rho}D^{(-)}_{0} \partial_{\mu}\partial_{\nu}D^{(-)}_{0}
\label{d2-2}
\eea
and we can prove that
\bea
d_{\{\mu\nu\rho\}}
= {1\over 2}~\left[ \partial_{\mu}\partial_{\nu}\partial_{\rho} - {1\over 6} 
( \eta_{\mu\nu}\partial_{\rho} + \eta_{\mu\rho}\partial_{\nu} +\eta_{\nu\rho}\partial_{\mu}) ~\square\right]d_{2}
\eea
\bea
f_{\{\mu\nu\}\rho} 
= {1\over 6}~\left[ \partial_{\mu}\partial_{\nu}\partial_{\rho} + {1\over 2} 
(\eta_{\mu\rho}\partial_{\nu} + \eta_{\nu\rho}\partial_{\mu} - \eta_{\mu\nu}\partial_{\rho}) ~\square\right] d_{2}.
\eea

We also have:
\be
f_{\mu} \equiv  \eta^{\nu\rho}~f_{\{\mu\nu\}\rho} = {1 \over 2}~\partial_{\mu} \square d_{2}
\ee

Finally we have distributions with four derivatives:
\bea
d_{\{\mu\nu\rho\sigma\}} \equiv D^{(+)}_{0} \partial_{\mu}\partial_{\nu}\partial_{\rho}\partial_{\sigma}D^{(+)}_{0}
- D^{(-)}_{0} \partial_{\mu}\partial_{\nu}\partial_{\rho}\partial_{\sigma}D^{(-)}_{0}
= a~\Bigl[ \partial_{\mu}\partial_{\nu}\partial_{\rho}\partial_{\sigma} 
\nonumber\\
- {1\over 8}~( \eta_{\mu\nu}\partial_{\rho}\partial_{\sigma} + \eta_{\mu\rho}\partial_{\nu}\partial_{\sigma} 
+ \eta_{\nu\rho}\partial_{\mu}\partial_{\sigma} + \eta_{\mu\sigma}\partial_{\nu}\partial_{\rho} 
+ \eta_{\nu\sigma}\partial_{\mu}\partial_{\rho} + \eta_{\rho\sigma}\partial_{\mu}\partial_{\nu} ) ~\square 
\nonumber\\
+ {1\over 48}~( \eta_{\mu\nu}\eta_{\rho\sigma} + \eta_{\mu\rho}\eta_{\nu\sigma}  + \eta_{\mu\sigma}\eta_{\nu\rho} )\square^{2} \Bigl] d_{2}
\label{d2-3}
\eea

\bea
f_{\{\mu\nu\}\{\rho\sigma\}} \equiv \partial_{\rho}\partial_{\sigma}D^{(+)}_{0} \partial_{\mu}\partial_{\nu}D^{(+)}_{0}
- \partial_{\rho}\partial_{\sigma}D^{(-)}_{0} \partial_{\mu}\partial_{\nu}D^{(-)}_{0}
= \Bigl[ \left(a - {1\over 3} \right)~ \partial_{\mu}\partial_{\nu}\partial_{\rho}\partial_{\sigma} 
\nonumber\\
- \left({a\over 8} - {1\over 12}\right)~( \eta_{\mu\rho}\partial_{\nu}\partial_{\sigma} + \eta_{\nu\rho}\partial_{\mu}\partial_{\sigma} 
+ \eta_{\mu\sigma}\partial_{\nu}\partial_{\rho} +  \eta_{\nu\sigma}\partial_{\mu}\partial_{\rho}) \square
- {a\over 8} ~(\eta_{\mu\nu}\partial_{\rho}\partial_{\sigma} +  \eta_{\rho\sigma}\partial_{\mu}\partial_{\nu} ) ~\square 
\nonumber\\
+ {a\over 48}~( \eta_{\mu\nu}\eta_{\rho\sigma} + \eta_{\mu\rho}\eta_{\nu\sigma}  + \eta_{\mu\sigma}\eta_{\nu\rho} )\square^{2} \Bigl] d_{2}
\label{d2-4}
\eea

\bea
f_{\{\mu\nu\rho\}\sigma} \equiv \partial_{\sigma}D^{(+)}_{0} \partial_{\mu}\partial_{\nu}\partial_{\rho}D^{(+)}_{0}
- \partial_{\sigma}D^{(-)}_{0} \partial_{\mu}\partial_{\nu}\partial_{\rho}D^{(-)}_{0}
= \Bigl[ \left({1\over 2} - a \right)~ \partial_{\mu}\partial_{\nu}\partial_{\rho}\partial_{\sigma} 
\nonumber\\
+ \left({a\over 8} - {1\over 12}\right)~( \eta_{\mu\nu}\partial_{\rho}\partial_{\sigma} + \eta_{\mu\rho}\partial_{\nu}\partial_{\sigma} 
+ \eta_{\nu\rho}\partial_{\mu}\partial_{\sigma} ) \square
+ {a\over 8} ~(\eta_{\mu\sigma}\partial_{\nu}\partial_{\rho} + \eta_{\nu\sigma}\partial_{\mu}\partial_{\rho}
+ \eta_{\rho\sigma}\partial_{\mu}\partial_{\nu} ) ~\square 
\nonumber\\
- {a\over 48}~( \eta_{\mu\nu}\eta_{\rho\sigma} + \eta_{\mu\rho}\eta_{\nu\sigma}  + \eta_{\mu\sigma}\eta_{\nu\rho} )\square^{2} \Bigl] d_{2}
\label{d2-5}
\eea
where
\be
a = {2 \over 5}.
\ee
We also have:
\be
f^{\prime}_{\mu\nu} \equiv \eta^{\rho\sigma}~f_{\{\mu\rho\}\{\nu\sigma\}} 
= {1 \over 6}~\Bigl( \partial_{\mu}\partial_{\nu} + {1 \over 2}~\eta_{\mu\nu} \square \Bigl) \square d_{2}
\ee
\be
f^{\prime} \equiv \eta^{\mu\rho}~\eta^{\nu\rho}~f_{\{\mu\nu\}\{\rho\sigma\}} = {1 \over 2}~\square^{2} d_{2}
\ee
\be
f^{\prime\prime}_{\mu\nu} \equiv \eta^{\rho\sigma}~f_{\{\mu\nu\rho\}\sigma} = 
{1 \over 3}~\Bigl( \partial_{\mu}\partial_{\nu} - {1 \over 4}~\eta_{\mu\nu} \square \Bigl) \square d_{2} 
\ee

For two-loop contributions we have
\be
D_{(2)}(A(x),B(y)) = \sum p_{j}(\partial) d_{3}(x - y)
\label{two-loops}
\ee
where we will need the basic distribution
\be
d_{3}(x) \equiv {1\over 12} [ D_{0}^{(+)}(x)^{3} - D_{0}^{(+)}(- x)^{3} ]
\ee
which is also with causal support and it can be causally split as above in
\be
d_{3} = d_{3}^{\rm adv} - d_{3}^{\rm ret}.
\ee

As in the one-loop case we encounter the associated causal distributions
\bea
d_{3}^{(1)}(x) = \p^{\mu}D_{0}^{(+)}(x) \p^{\nu}D_{0}^{(+)}(x) \partial_{\mu}\partial_{\nu}D_{0}^{(+)}(x) 
- \p^{\mu}D_{0}^{(+)}(x) \p^{\nu}D_{0}^{(+)}(- x) \partial_{\mu}\partial_{\nu}D_{0}^{(+)}(-x)
\nonumber\\
= {1 \over 4}~\square^{2}d_{3}
\nonumber \\
d_{3}^{(2)}(x) = D_{0}^{(+)}(x) \partial_{\mu}\p_{\nu}D_{0}^{(+)}(x)\p^{\mu} \partial^{\nu}D_{0}^{(+)}(x) 
- D_{0}^{(+)}(- x) \partial_{\mu}\p_{\nu}D_{0}^{(+)}(- x) \p^{\mu}\partial^{\nu}D_{0}^{(+)}(-x)
\nonumber\\
= {1 \over 2}~\square^{2}d_{3}
\eea
so in the end we have
\be
D_{(2)}(A(x),B(y)) \sim \square^{2} d_{3}(x - y).
\label{two-loops-1}
\ee
\newpage
\section{The Lagrangian for Massless Gravity\label{gravity}}

We have the following result \cite{cohomology2}:
\begin{thm}
Let $T$ a a relative cocycle, i.e. a Wick polynomial verifying (\ref{gau1}) 
which is tri-linear in the fields and is of canonical dimension
$
\omega(T) = 5
$
and ghost number
$
gh(T) = 0.
$
Then:
(i) $T$ is (relatively) cohomologous to a non-trivial cocycle of the form:
\be
T = \kappa~\sum_{j=1}^{9}~T_{j}
\ee

\bea
T_{1} \equiv - h_{\mu\nu}~\p^{\mu}h~\p^{\nu}h, \quad
T_{2} \equiv 2~h^{\mu\nu}~\p_{\mu}h_{\rho\sigma}~\p_{\nu}h^{\rho\sigma}, \quad 
T_{3} \equiv 4~h_{\mu\nu}~\p_{\rho}h^{\mu\sigma}~\p_{\sigma}h^{\nu\rho}
\nonumber \\
T_{4} \equiv 2~h_{\mu\nu}~\p_{\rho}h^{\mu\nu}~\p^{\rho}h, \quad
T_{5} \equiv - 4~h_{\mu\nu}~\p_{\rho}h^{\mu\sigma}~\p^{\rho}{h^{\nu}}_{\sigma}, \quad 
T_{6} \equiv - 4~u^{\mu}~\p_{\rho}\tilde{u}_{\sigma}~\p_{\mu}h^{\rho\sigma}
\nonumber \\
T_{7} \equiv 4~\p_{\nu}u^{\rho}~\p_{\mu}\tilde{u}_{\rho}~h^{\mu\nu}, \quad 
T_{8} \equiv - 4~\p^{\nu}u_{\nu}~\p_{\mu}\tilde{u}_{\nu}~h^{\mu\nu}, \quad
T_{9} \equiv 4~\p_{\mu}u^{\rho}~\p_{\rho}\tilde{u}_{\nu}~h^{\mu\nu}
\eea
where
$
\kappa \in \R.
$

(ii) The relation 
$
d_{Q}T = i~\partial_{\mu}T^{\mu}
$
is verified by:
\be
T^{\mu} = \kappa~\sum_{j=1}^{19}~T_{j}^{\mu}
\ee

\bea
T_{1}^{\mu} \equiv 4~u^{\lambda}~\p_{\lambda}h_{\rho\sigma}~\p^{\rho}h^{\mu\sigma}, \quad
T_{2}^{\mu} \equiv - 2~u^{\lambda}~\p_{\lambda}h_{\rho\sigma}~\p^{\mu}h^{\rho\sigma}, \quad 
T^{\mu}_{3} \equiv - 2~u^{\mu}~\p^{\lambda}h^{\rho\sigma}~\p_{\rho}h_{\lambda\sigma}
\nonumber \\
T_{4}^{\mu} \equiv - 4~\p_{\rho}u_{\nu}~\p_{\sigma}h^{\mu\nu}~h^{\rho\sigma}, \quad
T_{5}^{\mu} \equiv 4~\p_{\nu}u^{\nu}~\p^{\rho}h^{\mu\sigma}~h_{\rho\sigma}, \quad
T_{6}^{\mu} \equiv ~u^{\mu}~\p_{\lambda}h_{\rho\sigma}~\p^{\lambda}h^{\rho\sigma} 
\nonumber \\
T_{7}^{\mu} \equiv - 2~\p_{\nu}u^{\nu}~h_{\rho\sigma}~\p^{\mu}h^{\rho\sigma}, \quad
T_{8}^{\mu} \equiv - {1\over 2}~u^{\mu}~\p_{\nu}h~\p^{\nu}h, \quad
T_{9}^{\mu} \equiv \p^{\nu}u_{\nu}~h~\p^{\mu}h, \quad
\nonumber \\
T_{10}^{\mu} \equiv u^{\nu}~\p^{\mu}h~\p_{\nu}h, 
T_{11}^{\mu} \equiv - 2~\p_{\rho}u_{\sigma}~h^{\rho\sigma}~\p^{\mu}h, \quad
T_{12}^{\mu} \equiv 4~\p^{\rho}u_{\nu}~\p^{\mu}h^{\nu\sigma}~h_{\rho\sigma}, \quad
\nonumber \\
T_{13}^{\mu} \equiv - 4~\p^{\rho}u^{\nu}~\p_{\nu}h^{\mu\sigma}~h_{\rho\sigma}, \quad
T_{14}^{\mu} \equiv - 2~u^{\rho}~\p_{\rho}u_{\nu}~\p^{\mu}\tilde{u}^{\nu}
T_{15}^{\mu} \equiv 2~u^{\rho}~\p^{\sigma}u^{\mu}~\p_{\rho}\tilde{u}_{\sigma}, \quad
\nonumber \\
T_{16}^{\mu} \equiv - 2~u^{\mu}~\p_{\rho}u_{\sigma}~\p^{\sigma}\tilde{u}^{\rho}, \quad
T_{17}^{\mu} \equiv 2~\p^{\rho}u_{\rho}~\p^{\nu}u^{\mu}~\tilde{u}_{\nu}, \quad
T_{18}^{\mu} \equiv 2~u^{\rho}~\p_{\rho}\p_{\sigma}u^{\sigma}~\tilde{u}^{\mu}, \quad
\nonumber \\
T_{19}^{\mu} \equiv - 2~u^{\mu}~\p_{\rho}\p_{\sigma}u^{\sigma}~\tilde{u}^{\rho}
\eea

(iii) The relation 
$
d_{Q}T^{\mu} = i~\partial_{\nu}T^{\mu\nu}
$
is verified by:
\be
T^{\mu\nu} = \kappa~\sum_{j=1}^{5}~T_{j}^{\mu\nu}
\ee

\bea
T^{\alpha\beta}_{1} \equiv 2~u^{\mu}~\p_{\mu}u_{\nu}~\p^{\beta}h^{\alpha\nu}
- (\alpha \leftrightarrow \beta)
\nonumber \\
T^{\alpha\beta}_{2} \equiv 2~u^{\mu}~\p_{\nu}u^{\alpha}~\p_{\mu}h^{\beta\nu}
- (\alpha \leftrightarrow \beta)
\nonumber \\
T^{\alpha\beta}_{3} \equiv - 2~u^{\alpha}~\p_{\nu}u_{\mu}~\p^{\mu}h^{\beta\nu}
- (\alpha \leftrightarrow \beta)
\nonumber \\
T^{\alpha\beta}_{4} \equiv 4~\p^{\mu}u^{\alpha}~\p^{\nu}u^{\beta}~h_{\mu\nu}
\nonumber \\
T^{\alpha\beta}_{5} \equiv 2~\p_{\nu}u^{\nu}~\p_{\mu}u^{\alpha}~h^{\mu\beta}
- (\alpha \leftrightarrow \beta)
\eea

(iv) The relation 
$
d_{Q}T^{\mu\nu} = i~\partial_{\rho}T^{\mu\nu\rho}
$
is verified by:
\be
T^{\mu\nu\rho} = \kappa~\sum_{j=1}^{6}~T_{j}^{\mu\nu\rho}
\ee

\bea
T^{\alpha\beta\gamma}_{1} \equiv \p^{\alpha}u^{\beta}~u_{\mu}~\p^{\mu}u^{\gamma}
- (\alpha \leftrightarrow \beta)
\nonumber \\
T^{\alpha\beta\gamma}_{2} \equiv \p^{\beta}u^{\gamma}~u_{\mu}~\p^{\mu}u^{\alpha}
- (\alpha \leftrightarrow \beta)
\nonumber \\
T^{\alpha\beta\gamma}_{3} \equiv \p^{\gamma}u^{\alpha}~u_{\mu}~\p^{\mu}u^{\beta}
- (\alpha \leftrightarrow \beta)
\nonumber \\
T^{\alpha\beta\gamma}_{4} \equiv \p^{\beta}u^{\mu}~\p_{\mu}u^{\alpha}~u^{\gamma}
- (\alpha \leftrightarrow \beta)
\nonumber \\
T^{\alpha\beta\gamma}_{5} \equiv \p^{\gamma}u^{\mu}~\p_{\mu}u^{\beta}~u^{\alpha}
- (\alpha \leftrightarrow \beta)
\nonumber \\
T^{\alpha\beta\gamma}_{6} \equiv \p^{\alpha}u^{\mu}~\p_{\mu}u^{\gamma}~u^{\beta}
- (\alpha \leftrightarrow \beta)
\eea

and we have
$
d_{Q}T^{\mu\nu\rho} = 0.
$

(v) The cocycles
$
T, T^{\mu}, T^{\mu\nu}
$
and
$
T^{\mu\nu\rho}
$
are non-trivial and invariant with respect to parity. 
\label{T1gravity}
\end{thm}

In the preceding expressions we have, for simplicity, omitted the Wick ordering signs.
\newpage
\section{The Generic Expressions for the One-Loop Cochains}
We compute the one-loop contribution
$
D_{(1)}(T(x_{1}),T(x_{2})) = [ T(x_{1}),T(x_{2})]_{(1)}
$
using Wick theorem: we get
\bea
T_{j}(x_{1})T_{k}(x_{2}) = :T_{j}(x_{1})T_{k}(x_{2}):
\nonumber\\
+ \sum P_{j,0}(\p)D^{(+)}_{0}(x_{1} - x_{2}) W_{j,0}(x_{1},x_{2}) 
+ \sum P_{j,1}(\p)d^{(+)}_{2}(x_{1} - x_{2}) W_{j,1}(x_{1},x_{2})
\nonumber\\
+ \sum P_{j,2}(\p)d^{(+)}_{3}(x_{1} - x_{2}) W_{j,2}(x_{1},x_{2}),\qquad
j,k = 1, \dots, 9
\eea
where the last three terms correspond to tree, one-loop and two-loop contributions .
We obtain the following formula for the one-loop contribution:
\bea
D_{(1)}(T(x_{1}),T(x_{2})) = 
d_{\mu\nu}(x_{1} - x_{2})~A_{1}^{\{\mu\nu\}}(x_{1},x_{2})
+ f_{\mu\nu}(x_{1} - x_{2})~A_{2}^{\{\mu\nu\}}(x_{1},x_{2})
\nonumber\\
+ f(x_{1} - x_{2})~A_{3}(x_{1},x_{2})
+ f_{\{\rho\sigma\}\mu}(x_{1} - x_{2})~A_{4}^{\{\rho\sigma\}\mu}(x_{1},x_{2})
\nonumber\\
+ f^{\prime}_{\rho}(x_{1} - x_{2})~A_{5}^{\rho}(x_{1},x_{2})
+ f_{\{\mu\nu\}\{\rho\sigma\}}(x_{1} - x_{2})~A_{6}^{\{\mu\nu\}\{\rho\sigma\}}(x_{1},x_{2})
\nonumber\\
+ f^{\prime}_{\rho\sigma}(x_{1} - x_{2})~A_{7}^{\{\rho\sigma\}}(x_{1},x_{2})
+ f^{\prime}(x_{1} - x_{2})~A_{8}(x_{1},x_{2})
\label{D-one-loop-A}
\eea
where
\bea
A_{1}^{\{\mu\nu\}}(x_{1},x_{2}) = {\cal S}_{\mu\nu}~
[ - 24~\partial^{\mu}h(x_{1})~\partial^{\nu}h(x_{2})
+ 64~\partial^{\rho}h^{\mu\sigma}(x_{1})~\partial_{\rho}{h^{\nu}}_{\sigma}(x_{2})
\nonumber\\
+ 80~\partial^{\mu}h^{\rho\sigma}(x_{1})~\partial^{\nu}h_{\rho\sigma}(x_{2})
+ 32~\partial_{\rho}h^{\mu\rho}(x_{1})~\partial_{\sigma}h^{\nu\sigma}(x_{2}) ]
\nonumber\\
+ {\cal S}_{\mu\nu}~
\{ [ 8~\partial_{\rho}h(x_{1})~\partial^{\mu}h^{\nu\rho}(x_{2})
+ 24~\partial^{\mu}h(x_{1})~\partial_{\rho}h^{\nu\rho}(x_{2})
\nonumber\\
- 16~\partial^{\sigma}h_{\rho\sigma}(x_{1})~\partial^{\mu}h^{\nu\rho}(x_{2})
- 80~\partial^{\mu}h_{\rho\sigma}(x_{1})~\partial^{\rho}h^{\nu\sigma}(x_{2})
\nonumber\\
+ 16~\partial^{\sigma}h_{\rho\sigma}(x_{1})~\partial^{\rho}h^{\mu\nu}(x_{2})
\nonumber\\
+ 16~\partial^{\mu}\tilde{u}^{\rho}(x_{1})~\partial_{\rho}u^{\nu}(x_{2})
+ 16~\partial^{\rho}\tilde{u}^{\mu}(x_{1})~\partial_{\rho}u^{\nu}(x_{2})
\nonumber\\
- 32~\partial_{\rho}\tilde{u}^{\rho}(x_{1})~\partial^{\mu}u^{\nu}(x_{2})
- 16~\partial^{\mu}\tilde{u}^{\nu}(x_{1})~\partial_{\rho}u^{\rho}(x_{2}) ]
\nonumber\\
+ ( x_{1} \leftrightarrow x_{2} ) \}
\eea
\bea
A_{2}^{\{\mu\nu\}}(x_{1},x_{2}) = {\cal S}_{\mu\nu}~
[ - 20~\partial^{\mu}h(x_{1})~\partial^{\nu}h(x_{2})
+ 64~\partial_{\rho}h^{\mu\sigma}(x_{1})~\partial_{\sigma}h^{\nu\rho}(x_{2})
\nonumber\\
+ 80~\partial^{\mu}h^{\rho\sigma}(x_{1})~\partial^{\nu}h_{\rho\sigma}(x_{2})
+ 32~\partial_{\rho}h^{\mu\rho}(x_{1})~\partial_{\sigma}h^{\nu\sigma}(x_{2}) ]
\nonumber\\
+ {\cal S}_{\mu\nu}~
\{ [ 24~\partial^{\mu}h(x_{1})~\partial_{\rho}h^{\nu\rho}(x_{2})
- 80~\partial^{\mu}h_{\rho\sigma}(x_{1})~\partial^{\rho}h^{\nu\sigma}(x_{2})
\nonumber\\
+ 16~\partial^{\mu}\tilde{u}^{\rho}(x_{1})~\partial_{\rho}u^{\nu}(x_{2})
+ 16~\partial^{\rho}\tilde{u}^{\mu}(x_{1})~\partial_{\rho}u^{\nu}(x_{2})
\nonumber\\
- 16~\partial_{\rho}\tilde{u}^{\rho}(x_{1})~\partial^{\mu}u^{\nu}(x_{2})
- 16~\partial^{\mu}\tilde{u}^{\nu}(x_{1})~\partial_{\rho}u^{\rho}(x_{2}) ]
\nonumber\\
+ ( x_{1} \leftrightarrow x_{2} ) \}
\eea
\be
A_{3}(x_{1},x_{2}) = 8~\partial_{\rho}\tilde{u}^{\rho}(x_{1})~\partial_{\sigma}u^{\sigma}(x_{2})
+ ( x_{1} \leftrightarrow x_{2} )
\ee
\bea
A_{4}^{\{\rho\sigma\}\mu}(x_{1},x_{2}) = {\cal S}_{\rho\sigma}~
\{ [ - 8~h^{\mu\rho}(x_{1})~\partial^{\sigma}h(x_{2})
+ 112~h^{\mu\rho}(x_{1})~\partial_{\nu}h^{\nu\sigma}(x_{2})
\nonumber\\ 
+ 16~{h^{\rho}}_{\nu}(x_{1})~\partial^{\sigma}h^{\mu\nu}(x_{2})
- 16~{h^{\mu}}_{\nu}(x_{1})~\partial^{\rho}h^{\nu\sigma}(x_{2})
\nonumber\\
+ 16~h(x_{1})~\partial^{\mu}h^{\rho\sigma}(x_{2})
- 16~h(x_{1})~\partial^{\rho}h^{\mu\sigma}(x_{2})
\nonumber\\
- 16~h^{\mu\nu}(x_{1})~\partial_{\nu}h^{\rho\sigma}(x_{2})
\nonumber\\
+ 8~u^{\rho}(x_{1})~\partial^{\mu}\tilde{u}^{\sigma}(x_{2})
+ 8~u^{\rho}(x_{1})~\partial^{\sigma}\tilde{u}^{\mu}(x_{2})
\nonumber\\
+ 16~u^{\mu}(x_{1})~\partial^{\rho}\tilde{u}^{\sigma}(x_{2})
]
\nonumber\\
- ( x_{1} \leftrightarrow x_{2} ) \}
\eea
\bea
A_{5}^{\rho}(x_{1},x_{2}) = 
\{ [ 8~h^{\rho\sigma}(x_{1})~\partial_{\sigma}h(x_{2})
- 8~h(x_{1})~\partial^{\rho}h(x_{2})
\nonumber\\ 
+ 56~h_{\alpha\beta}(x_{1})~\partial^{\rho}h^{\alpha\beta}(x_{2})
- 16~h^{\rho\sigma}(x_{1})~\partial^{\nu}h_{\nu\sigma}(x_{2})
\nonumber\\
- 64~h_{\nu\sigma}(x_{1})~\partial^{\nu}h^{\rho\sigma}(x_{2})
- 16~h(x_{1})~\partial_{\sigma}h^{\rho\sigma}(x_{2})
\nonumber\\
- 8~u^{\rho}(x_{1})~\partial_{\sigma}\tilde{u}^{\sigma}(x_{2})
]
\nonumber\\
- ( x_{1} \leftrightarrow x_{2} ) \}
\eea
\be
A_{6}^{\{\mu\nu\}\{\rho\sigma\}\mu}(x_{1},x_{2}) = - 48~{\cal S}_{\mu\nu}~{\cal S}_{\rho\sigma}~
[ h^{\mu\rho}(x_{1})~h^{\nu\sigma}(x_{2}) ] 
+ 32~h^{\mu\nu}(x_{1})~h^{\rho\sigma}(x_{2}) 
\ee
\be
A_{7}^{\{\rho\sigma\}\mu}(x_{1},x_{2}) = 40~
[ h^{\rho\sigma}(x_{1})~h(x_{2}) + h(x_{1})~h^{\rho\sigma}(x_{2}) ]
\ee
\be
A_{8}(x_{1},x_{2}) = - 16~h_{\rho\sigma}(x_{1})~h^{\rho\sigma}(x_{2}) - 12~h(x_{1})~h(x_{2})
\ee
Using the formulas from the preceding Section one can rewrite everything more compactly:
\bea
D_{(1)}(T(x_{1}),T(x_{2})) = 
\partial_{\mu}\partial_{\nu}d_{2}(x_{1} - x_{2})~a_{1}^{\mu\nu}(x_{1},x_{2})
+ \square d_{2}(x_{1} - x_{2})~a_{2}(x_{1},x_{2})
\nonumber\\
+ \partial_{\mu}\partial_{\rho}\partial_{\sigma}d_{2}(x_{1} - x_{2})~a_{3}^{\{\mu\rho\sigma\}}(x_{1},x_{2})
+ \partial_{\mu}\square d_{2}(x_{1} - x_{2})~a_{4}^{\mu}(x_{1},x_{2})
\nonumber\\
+ \partial_{\mu}\partial_{\nu}\partial_{\rho}\partial_{\sigma}d_{2}(x_{1} - x_{2})~a_{5}^{\{\mu\nu\rho\sigma\}}(x_{1},x_{2})
+ \partial_{\mu}\partial_{\nu}\square d_{2}(x_{1} - x_{2})~a_{6}^{\{\mu\nu\}}(x_{1},x_{2})
\nonumber\\
+ \square^{2} d_{2}(x_{1} - x_{2})~a_{7}(x_{1},x_{2})
\label{D-one-loop-a}
\eea
where
\bea
a_{1}^{\{\mu\nu\}}(x_{1},x_{2}) = {1 \over 3}~{\cal S}_{\mu\nu}~
[ - 68~\partial^{\mu}h(x_{1})~\partial^{\nu}h(x_{2})
+ 128~\partial^{\rho}h^{\mu\sigma}(x_{1})~\partial_{\rho}{h^{\nu}}_{\sigma}(x_{2})
\nonumber\\
+ 240~\partial^{\mu}h^{\rho\sigma}(x_{1})~\partial^{\nu}h_{\rho\sigma}(x_{2})
+ 96~\partial_{\rho}h^{\mu\rho}(x_{1})~\partial_{\sigma}h^{\nu\sigma}(x_{2}) 
+ 64~\partial_{\rho}h^{\mu\sigma}(x_{1})~\partial_{\sigma}h^{\nu\rho}(x_{2}) ]
\nonumber\\
+ {1 \over 3}~{\cal S}_{\mu\nu}~
\{ [ 72~\partial^{\mu}h(x_{1})~\partial_{\rho}h^{\nu\rho}(x_{2})
- 32~\partial^{\sigma}h_{\rho\sigma}(x_{1})~\partial^{\mu}h^{\nu\sigma}(x_{2})
\nonumber\\
- 240~\partial^{\mu}h_{\rho\sigma}(x_{1})~\partial^{\rho}h^{\nu\sigma}(x_{2})
+ 32~\partial^{\sigma}h_{\rho\sigma}(x_{1})~\partial^{\rho}h^{\mu\nu}(x_{2})
\nonumber\\
+ 32~\partial_{\rho}h(x_{1})~\partial^{\mu}h^{\nu\rho}(x_{2})
\nonumber\\
+ 48~\partial^{\mu}\tilde{u}^{\rho}(x_{1})~\partial_{\rho}u^{\nu}(x_{2}) 
+ 48~\partial^{\rho}\tilde{u}^{\mu}(x_{1})~\partial_{\rho}u^{\nu}(x_{2}) 
\nonumber\\
- 80~\partial_{\rho}\tilde{u}^{\rho}(x_{1})~\partial^{\mu}u^{\nu}(x_{2}) 
- 48~\partial^{\mu}\tilde{u}^{\nu}(x_{1})~\partial_{\rho}u^{\rho}(x_{2}) ]
\nonumber\\
+ ( x_{1} \leftrightarrow x_{2}) \}
\label{a1}
\eea
\bea
a_{2}(x_{1},x_{2}) = {1 \over 3}~
[ 2~\partial^{\mu}h(x_{1})~\partial_{\mu}h(x_{2})
- 32~\partial_{\mu}h_{\rho\sigma}(x_{1})~\partial^{\mu}h^{\rho\sigma}(x_{2})
\nonumber\\
+ 32~\partial_{\mu}h_{\rho\sigma}(x_{1})~\partial^{\rho}h^{\mu\sigma}(x_{2})
+ 16~\partial_{\rho}h^{\mu\rho}(x_{1})~\partial^{\sigma}h_{\mu\sigma}(x_{2}) ]
\nonumber\\
+ {1 \over 3}~
\{ [ - 12~\partial^{\mu}h(x_{1})~\partial^{\nu}h_{\mu\nu}(x_{2})
+ 32~\partial_{\mu}\tilde{u}^{\mu}(x_{1})~\partial_{\rho}u^{\rho}(x_{2}) ]
\nonumber\\
+ ( x_{1} \leftrightarrow x_{2}) \}
\label{a2}
\eea
\bea
a_{3}^{\{\mu\rho\sigma\}}(x_{1},x_{2}) = {1 \over 3}~{\cal S}_{\mu\rho\sigma}~
\{ [ - 4~h^{\rho\sigma}(x_{1})~\partial^{\mu}h(x_{2})
+ 56~h^{\rho\sigma}(x_{1})~\partial_{\nu}h^{\mu\nu}(x_{2})
\nonumber\\
- 8~h^{\mu\nu}(x_{1})~\partial_{\nu}h^{\rho\sigma}(x_{2})
+ 16~u^{\mu}(x_{1})~\partial^{\rho}\tilde{u}^{\sigma}(x_{2}) ]
\nonumber\\
- ( x_{1} \leftrightarrow x_{2}) \}
\label{a3}
\eea
\bea
a_{4}^{\mu}(x_{1},x_{2}) = {1 \over 3}~
\{ [ 16~h^{\mu\nu}(x_{1})~\partial_{\nu}h(x_{2})
- 22~h(x_{1})~\partial^{\mu}h(x_{2})
\nonumber\\
- 16~h^{\mu\nu}(x_{1})~\partial^{\rho}h_{\nu\rho}(x_{2})
+ 12~h(x_{1})~\partial_{\nu}h^{\mu\nu}(x_{2})
\nonumber\\
+ 84~h_{\rho\sigma}(x_{1})~\partial^{\mu}h^{\rho\sigma}(x_{2})
- 112~h_{\rho\sigma}(x_{1})~\partial^{\rho}h^{\mu\sigma}(x_{2})
\nonumber\\
- 12~u^{\mu}(x_{1})~\partial_{\nu}\tilde{u}^{\nu}(x_{2}) 
+ 4~u_{\nu}(x_{1})~\partial^{\nu}\tilde{u}^{\mu}(x_{2}) 
+ 4~u_{\nu}(x_{1})~\partial^{\mu}\tilde{u}^{\nu}(x_{2}) ]
]
\nonumber\\
- ( x_{1} \leftrightarrow x_{2}) \}
\label{a4}
\eea
\be
a_{5}^{\{\mu\nu\rho\sigma\}}(x_{1},x_{2}) = 
- 16~\Bigl( a - {1 \over 3} \Bigl)~{\cal S}_{\mu\nu\rho\sigma}~[ h^{\mu\nu}(x_{1})~h^{\rho\sigma}(x_{2}) ]
\label{a5}
\ee
\bea
a_{6}^{\{\mu\nu\}}(x_{1},x_{2}) = 
\Bigl( 2 a + {8 \over 3} \Bigl)~[ h^{\mu\nu}(x_{1})~h(x_{2}) + h(x_{1})~h^{\mu\nu}(x_{2})]
\nonumber\\
+ 8~\Bigl( a + {1 \over 3} \Bigl){\cal S}_{\mu\nu}~[h^{\mu\rho}(x_{1})~{h^{\nu}}_{\rho}(x_{2})]
\label{a6}
\eea
\be
a_{7}(x_{1},x_{2}) = {1 \over 3}~( - a + 2 )~h(x_{1})~h(x_{2})
- \Bigl( { 2 a \over 3}  + 8 \Bigl)~h^{\mu\nu}(x_{1})~h_{\mu\nu}(x_{2})
\label{a7}
\ee

The expressions
$
D_{(1)}(T^{I}(x_{1}),T^{J}(x_{2})), |I| + |J| =1, 2
$
can be computed in the same way.
As far as we know, the expressions (\ref{D-one-loop-a}) $+$ (\ref{a1}) - (\ref{a7}) are new in the literature.

\newpage

\section{Cohomology}
In this Section we prove the main result:
\begin{thm}
The expression (\ref{D-one-loop-a}) is in fact a relative coboundary, i.e. of the form:
\bea
D_{(1)}(T(x_{1}),T(x_{2})) = d_{Q}B(x_{1},x_{2}) + i~(\delta B)(x_{1},x_{2})
\nonumber\\
= d_{Q}B(x_{1},x_{2}) + i~[ \p_{\mu}^{1}B^{\mu}(x_{1},x_{2}) - (1 \leftrightarrow 2)]
\eea
where the generic form of the cochains is:
\be
C(x_{1},x_{2}) = \sum_{j} p_{j}(\p) d_{2}(x_{1} - x_{2})~W_{j}(x_{1},x_{2})
\label{cochain}
\ee
with 
$
W_{j}
$ 
Wick monomials. The same assertion stays true for one-loop the chronological product
$
T_{(1)}(T(x_{1}),T(x_{2}))
$.
\end{thm}

{\bf Proof:}
The proof is very computational: we must make a generic ansatz of the form (\ref{cochain})  for the expressions 
$B(x_{1},x_{2})$ 
and
$
B^{\mu}(x_{1},x_{2})
$.
The expression $B$ should be antisymmetric with respect to
$1 \leftrightarrow 2$
as it is 
$
D_{(1)}(T(x_{1}),T(x_{2}))
$.

After some hard computations we can find a non-trivial solution of this problem. It is more interesting to emphasize that every
individual term from (\ref{D-one-loop-a}) is in fact a coboundary. We have $38$ terms in (\ref{D-one-loop-a}):

1) 
$
D = \p_{\mu}\p_{\nu}\square d_{2}(x_{1} - x_{2})~[ h^{\mu\nu}(x_{1}) h(x_{2}) + (1 \leftrightarrow 2) ]
$

We can take
$
B = 0
$
and
\be
B^{\mu}(x_{1},x_{2}) = - i~\p^{\mu}\p^{\nu}\p^{\rho}d_{2}(x_{1} - x_{2})~h_{\nu\rho}(x_{1}) h(x_{2})
+ i \p^{\nu}\p^{\rho}d_{2}(x_{1} - x_{2})~\p^{\mu}h_{\nu\rho}(x_{1}) h(x_{2})
\ee

2) 
$
D = \p_{\mu}\p_{\nu}\square d_{2}(x_{1} - x_{2})~h^{\mu\rho}(x_{1})~{h^{\nu}}_{\rho}(x_{2})
$

We can take
$
B = 0
$
and
\be
B^{\mu}(x_{1},x_{2}) = - {i \over 2}[~\p^{\mu}\p^{\nu}\p_{\rho}d_{2}(x_{1} - x_{2})~h_{\nu\sigma}(x_{1}) h^{\rho\sigma}(x_{2})
- \p^{\nu}\p_{\rho}d_{2}(x_{1} - x_{2})~\p^{\mu}h_{\nu\sigma}(x_{1}) h^{\rho\sigma}(x_{2}) ]
\ee

3) 
$
D = \square^{2} d_{2}(x_{1} - x_{2})~h^{\mu\nu}(x_{1}) h_{\mu\nu}(x_{2})
$

We can take
$
B = 0
$
and
\be
B^{\mu}(x_{1},x_{2}) = - {i\over 2}~[\p^{\mu}\square d_{2}(x_{1} - x_{2})~h_{\nu\rho}(x_{1}) h^{\nu\rho}(x_{2})
- \square d_{2}(x_{1} - x_{2})~\p^{\mu}h_{\nu\rho}(x_{1}) h^{\nu\rho}(x_{2})]
\ee

4) 
$
D = \square^{2} d_{2}(x_{1} - x_{2})~h(x_{1}) h(x_{2})
$

We can take
$
B = 0
$
and
\be
B^{\mu}(x_{1},x_{2}) = - {i\over 2}~[\p^{\mu}\square d_{2}(x_{1} - x_{2})~h(x_{1}) h(x_{2})
- \square d_{2}(x_{1} - x_{2})~\p^{\mu}h(x_{1}) h(x_{2})]
\ee

\newpage
5) 
$
D = \p_{\mu}\p_{\nu}\p_{\rho}\p_{\sigma}d_{2}(x_{1} - x_{2})~h^{\mu\nu}(x_{1}) h^{\rho\sigma}(x_{2})
$

We can take
\be
B(x_{1},x_{2})  = {i\over 2}~\p^{\mu}\p^{\nu}d_{2}(x_{1} - x_{2})~[ \p^{\rho}h_{\mu\rho}(x_{1}) \tilde{u}_{\nu}(x_{2}) + (1 \leftrightarrow 2)]
\ee
and
\be
B^{\mu}(x_{1},x_{2}) = - {i\over 2}~[\p_{\nu}\p_{\rho}\p_{\sigma}d_{2}(x_{1} - x_{2})~h^{\mu\nu}(x_{1}) h^{\rho\sigma}(x_{2})
+ \p_{\rho}\p_{\sigma}d_{2}(x_{1} - x_{2})~h^{\mu\nu}(x_{1}) \p_{\sigma}h^{\rho\sigma}(x_{2})]
\ee

6) 
$
D = \p_{\mu}\p_{\rho}\p_{\sigma}d_{2}(x_{1} - x_{2})~[ h^{\rho\sigma}(x_{1}) \p^{\mu}h(x_{2}) - (1 \leftrightarrow 2)]
$

We can take
$
B = 0
$
and
\be
B^{\mu}(x_{1},x_{2}) = i~\p_{\nu}\p_{\rho}d_{2}(x_{1} - x_{2})~\p^{\mu}h(x_{1}) h^{\nu\rho}(x_{2})
\ee

7) 
$
D = \p_{\mu}\p_{\rho}\p_{\sigma}d_{2}(x_{1} - x_{2})~[ h^{\rho\sigma}(x_{1}) \p_{\nu}h^{\mu\nu}(x_{2}) - (1 \leftrightarrow 2)]
$

We can take
\be
B(x_{1},x_{2}) = i~\p^{\mu}\p^{\nu}d_{2}(x_{1} - x_{2})~[ \p^{\rho}h_{\mu\rho}(x_{1}) \tilde{u}_{\nu}(x_{2}) + (1 \leftrightarrow 2)]
\ee
and
\be
B^{\mu}(x_{1},x_{2}) = - i~\p_{\nu}\p_{\rho}d_{2}(x_{1} - x_{2})~h^{\mu\nu}(x_{1}) \p_{\sigma}h^{\rho\sigma}(x_{2})
\ee

8) 
$
D = \p_{\mu}\p_{\rho}\p_{\sigma}d_{2}(x_{1} - x_{2})~[ h^{\mu\nu}(x_{1}) \p_{\nu}h^{\rho\sigma}(x_{2}) - (1 \leftrightarrow 2)]
$

We can take
\be
B(x_{1},x_{2}) = i~\p^{\mu}d_{2}(x_{1} - x_{2})~[ \p_{\rho}h^{\nu\rho}(x_{1}) \p_{\nu}\tilde{u}_{\mu}(x_{2}) - (1 \leftrightarrow 2)]
\ee
and
\be
B^{\mu}(x_{1},x_{2}) = i~\p_{\nu}\p_{\rho}d_{2}(x_{1} - x_{2})~\p_{\sigma}h^{\mu\nu}(x_{1}) h^{\rho\sigma}(x_{2})
- i~\p_{\nu}d_{2}(x_{1} - x_{2})~h^{\mu\sigma}(x_{1}) \p_{\rho}\p_{\sigma}h^{\nu\rho}(x_{2})
\ee

9) 
$
D = \p_{\mu}\p_{\rho}\p_{\sigma}d_{2}(x_{1} - x_{2})~[ u^{\mu}(x_{1}) \p^{\rho} \tilde{u}^{\sigma}(x_{2}) - (1 \leftrightarrow 2)]
$

We can take
$
B = 0
$
and
\be
B^{\mu}(x_{1},x_{2}) = - i~\p_{\nu}\p_{\rho}d_{2}(x_{1} - x_{2})~\p^{\mu}\tilde{u}^{\nu}(x_{1}) u^{\rho}(x_{2})
\ee

10) 
$
D = \p_{\mu}\square d_{2}(x_{1} - x_{2})~[ h^{\mu\nu}(x_{1}) \p^{\rho}h_{\nu\rho}(x_{2}) - (1 \leftrightarrow 2)]
$

We can take
$
B = 0
$
and
\be
B^{\mu}(x_{1},x_{2}) = i~\p^{\mu}\p^{\nu}d_{2}(x_{1} - x_{2})~\p_{\sigma}h^{\rho\sigma}(x_{1}) h_{\nu\rho}(x_{2})
- i~\p_{\nu}d_{2}(x_{1} - x_{2})~\p^{\mu}\p^{\sigma}h_{\rho\sigma}(x_{1}) h^{\nu\rho}(x_{2})
\ee

11) 
$
D = \p_{\mu}\square d_{2}(x_{1} - x_{2})~[ h(x_{1}) \p_{\nu}h^{\mu\nu}(x_{2}) - (1 \leftrightarrow 2)]
$

We can take
$
B = 0
$
and
\be
B^{\mu}(x_{1},x_{2}) = i~\p^{\mu}\p^{\nu}d_{2}(x_{1} - x_{2})~\p^{\rho}h_{\nu\rho}(x_{1}) h(x_{2})
- i~\p^{\nu}d_{2}(x_{1} - x_{2})~\p^{\mu}\p^{\rho}h_{\nu\rho}(x_{1}) h(x_{2})
\ee

\newpage

12) 
$
D = \p_{\mu}\square d_{2}(x_{1} - x_{2})~[ h_{\rho\sigma}(x_{1}) \p^{\mu}h^{\rho\sigma}(x_{2}) - (1 \leftrightarrow 2)]
$

We can take
$
B = 0
$
and
\be
B^{\mu}(x_{1},x_{2}) = i~\square d_{2}(x_{1} - x_{2})~\p^{\mu}h^{\nu\rho}(x_{1}) h_{\nu\rho}(x_{2})
\ee

13) 
$
D = \p_{\mu}\square d_{2}(x_{1} - x_{2})~[ h(x_{1}) \p^{\mu}h(x_{2}) - (1 \leftrightarrow 2)]
$

We can take
$
B = 0
$
and
\be
B^{\mu}(x_{1},x_{2}) = i~\square d_{2}(x_{1} - x_{2})~\p^{\mu}h(x_{1}) h(x_{2})
\ee

14) 
$
D = \p_{\mu}\square d_{2}(x_{1} - x_{2})~[ h^{\mu\nu}(x_{1}) \p_{\nu}h(x_{2}) - (1 \leftrightarrow 2)]
$

We can take
$
B = 0
$
and
\be
B^{\mu}(x_{1},x_{2}) = i~\p^{\mu}\p^{\nu} d_{2}(x_{1} - x_{2})~\p^{\rho}h(x_{1}) h_{\nu\rho}(x_{2})
- i~\p^{\nu} d_{2}(x_{1} - x_{2})~\p^{\mu}\p^{\rho}h(x_{1}) h_{\nu\rho}(x_{2})
\ee

15) 
$
D = \p_{\mu}\square d_{2}(x_{1} - x_{2})~[ h_{\rho\sigma}(x_{1}) \p^{\rho}h^{\mu\sigma}(x_{2}) - (1 \leftrightarrow 2)]
$

We can take
$
B = 0
$
and
\be
B^{\mu}(x_{1},x_{2}) = i~\p^{\mu}\p^{\nu} d_{2}(x_{1} - x_{2})~\p_{\rho}h_{\nu\sigma}(x_{1}) h^{\rho\sigma}(x_{2})
- i~\p_{\nu} d_{2}(x_{1} - x_{2})~\p^{\mu}\p^{\rho}h^{\nu\sigma}(x_{1}) h_{\rho\sigma}(x_{2})
\ee

16) 
$
D = \p_{\mu}\square d_{2}(x_{1} - x_{2})~[ u_{\nu}(x_{1}) \p^{\mu}\tilde{u}^{\nu}(x_{2}) - (1 \leftrightarrow 2)]
$

We can take
$
B = 0
$
and
\be
B^{\mu}(x_{1},x_{2}) = - i~\square d_{2}(x_{1} - x_{2})~\p^{\mu}\tilde{u}^{\nu}(x_{1}) u_{\nu}(x_{2})
\ee

17) 
$
D = \p_{\mu}\square d_{2}(x_{1} - x_{2})~[ u^{\mu}(x_{1}) \p_{\nu}\tilde{u}^{\nu}(x_{2}) - (1 \leftrightarrow 2)]
$

We can take
$
B = 0
$
and
\bea
B^{\mu}(x_{1},x_{2}) = - i~\p^{\mu}\p^{\nu}\p^{\rho}d_{2}(x_{1} - x_{2})~u_{\nu}(x_{1}) \tilde{u}_{\rho}(x_{2})
\nonumber\\
- i \p_{\nu}\square d_{2}(x_{1} - x_{2}) \tilde{u}^{\mu}(x_{1}) u^{\nu}(x_{2})
+ i \p_{\nu}\p_{\rho}d_{2}(x_{1} - x_{2})~\p^{\mu}u^{\nu}(x_{1}) \tilde{u}^{\rho}(x_{2})
\eea

18) 
$
D = \p_{\mu}\square d_{2}(x_{1} - x_{2})~[ u_{\nu}(x_{1}) \p^{\nu}\tilde{u}^{\mu}(x_{2}) - (1 \leftrightarrow 2)]
$

We can take
$
B = 0
$
and
\bea
B^{\mu}(x_{1},x_{2}) = - i~\p^{\mu}\p^{\nu}\p^{\rho}d_{2}(x_{1} - x_{2})~u_{\nu}(x_{1}) \tilde{u}_{\rho}(x_{2})
\nonumber\\
- i \p_{\nu}\square d_{2}(x_{1} - x_{2}) \tilde{u}^{\nu}(x_{1}) u^{\mu}(x_{2})
+ i \p_{\nu}\p_{\rho}d_{2}(x_{1} - x_{2})~\p^{\mu}u^{\nu}(x_{1}) \tilde{u}^{\rho}(x_{2})
\eea

19) 
$
D = \p_{\mu}\p_{\nu}d_{2}(x_{1} - x_{2})~[ \p^{\mu}h^{\nu\rho}(x_{1}) \p^{\sigma} h_{\rho\sigma}(x_{2}) + (1 \leftrightarrow 2)]
$

We can take
$
B = 0
$
and
\be
B^{\mu}(x_{1},x_{2}) = - i~\p_{\nu}d_{2}(x_{1} - x_{2})~\p^{\mu}h^{\nu\rho}(x_{1}) \p^{\sigma} h_{\rho\sigma}(x_{2})
\ee

\newpage
20) 
$
D = \p_{\mu}\p_{\nu}d_{2}(x_{1} - x_{2})~[ \p^{\rho}h^{\mu\nu}(x_{1}) \p^{\sigma}h_{\rho\sigma}(x_{2}) + (1 \leftrightarrow 2)]
$

We can take
\be
B(x_{1},x_{2}) = - i~d_{2}(x_{1} - x_{2})~[ \p_{\mu}\p_{\nu}h^{\nu\rho}(x_{1}) \p_{\rho}\tilde{u}^{\mu}(x_{2})
+ (1 \leftrightarrow 2)]
\ee
and
\be
B^{\mu}(x_{1},x_{2}) = - i~ \p^{\nu}d_{2}(x_{1} - x_{2})~\p^{\rho}h^{\mu\nu}(x_{1}) \p^{\sigma}h_{\rho\sigma}(x_{2})
\ee

21) 
$
D = \p_{\mu}\p_{\nu}d_{2}(x_{1} - x_{2})~\p_{\rho}h^{\mu\rho}(x_{1}) \p_{\sigma}h^{\nu\sigma}(x_{2})
$

We can take
\be
B(x_{1},x_{2}) = - {i \over 2}~\p^{\mu}\p^{\nu}d_{2}(x_{1} - x_{2})~[ \p^{\rho}h_{\mu\rho}(x_{1}) \tilde{u}_{\nu}(x_{2}) + (1 \leftrightarrow 2)]
\ee
and
$
B^{\mu} = 0.
$

22) 
$
D = \p_{\mu}\p_{\nu}d_{2}(x_{1} - x_{2})~[ \p^{\mu}h(x_{1}) \p_{\rho}h^{\nu\sigma}(x_{2}) + (1 \leftrightarrow 2)]
$

We can take
$
B = 0
$
and
\be
B^{\mu}(x_{1},x_{2}) = - i~\p^{\nu}d_{2}(x_{1} - x_{2})~\p^{\mu}h(x_{1}) \p^{\rho}h_{\nu\sigma}(x_{2})
\ee

23) 
$
D = \p_{\mu}\p_{\nu}d_{2}(x_{1} - x_{2})~\p^{\mu}h^{\rho\sigma}(x_{1}) \p^{\nu}h_{\rho\sigma}(x_{2})
$

We can take
$
B = 0
$
and
\be
B^{\mu}(x_{1},x_{2}) = - {i \over 2}~\p^{\nu}d_{2}(x_{1} - x_{2})~\p^{\mu}h^{\rho\sigma}(x_{1}) \p_{\nu}h_{\rho\sigma}(x_{2})
\ee

24) 
$
D = \p_{\mu}\p_{\nu}d_{2}(x_{1} - x_{2})~\p^{\mu}h(x_{1}) \p^{\nu}h(x_{2})
$

We can take
$
B = 0
$
and
\be
B^{\mu}(x_{1},x_{2}) = - {i \over 2}~\p^{\nu}d_{2}(x_{1} - x_{2})~\p^{\mu}h(x_{1}) \p_{\nu}h(x_{2})
\ee

25) 
$
D = \p_{\mu}\p_{\nu}d_{2}(x_{1} - x_{2})~[ \p^{\mu}h^{\nu\rho}(x_{1}) \p_{\rho}h(x_{2}) + (1 \leftrightarrow 2)]
$

We can take
$
B = 0
$
and
\be
B^{\mu}(x_{1},x_{2}) = - {i \over 2}~\p_{\nu}d_{2}(x_{1} - x_{2})~\p^{\mu}h^{\nu\rho}(x_{1}) \p_{\rho}h(x_{2})
\ee

26) 
$
D = \p_{\mu}\p_{\nu}d_{2}(x_{1} - x_{2})~[ \p^{\mu}h_{\rho\sigma}(x_{1}) \p^{\rho}h^{\nu\sigma}(x_{2}) + (1 \leftrightarrow 2)]
$

We can take
$
B = 0
$
and
\be
B^{\mu}(x_{1},x_{2}) = - i~\p^{\nu}d_{2}(x_{1} - x_{2})~\p^{\mu}h^{\rho\sigma}(x_{1}) \p_{\rho}h_{\nu\sigma}(x_{2})
\ee

27) 
$
D = \p_{\mu}\p^{\nu}d_{2}(x_{1} - x_{2})~\p^{\rho}h^{\mu\sigma}(x_{1}) \p_{\rho}h_{\nu\sigma}(x_{2})
$

We can take
$
B = 0
$
and
\be
B^{\mu}(x_{1},x_{2}) = - {i \over 2}~\p_{\nu}\p^{\rho}d_{2}(x_{1} - x_{2})~[ \p^{\mu}h^{\nu\sigma}(x_{1}) h_{\rho\sigma}(x_{2})
+ h^{\nu\sigma}(x_{1}) \p^{\mu}h_{\rho\sigma}(x_{2}) ]
\ee

\newpage
28) 
$
D = \p_{\mu}\p_{\nu}d_{2}(x_{1} - x_{2})~\p_{\rho}h^{\mu\sigma}(x_{1}) \p_{\sigma}h^{\nu\rho}(x_{2})
$

We can take
\be
B(x_{1},x_{2}) = {i \over 2}~d_{2}(x_{1} - x_{2})~[ \p_{\mu}\p_{\nu}h^{\nu\rho}(x_{1}) \p_{\rho}\tilde{u}^{\mu}(x_{2})
+ (1 \leftrightarrow 2)]
\ee
and
\be
B^{\mu}(x_{1},x_{2}) = - {i \over 2}~[ \p^{\nu}d_{2}(x_{1} - x_{2})~\p^{\rho}h^{\mu\sigma}(x_{1}) \p_{\sigma}h_{\nu\rho}(x_{2})
+ d_{2}(x_{1} - x_{2})~\p^{\rho}h^{\mu\sigma}(x_{1}) \p^{\nu}\p_{\sigma}h_{\nu\rho}(x_{2}) ]
\ee

29) 
$
D = \p_{\mu}\p_{\nu}d_{2}(x_{1} - x_{2})~[ \p^{\rho}u^{\mu}(x_{1}) \p^{\nu}\tilde{u}_{\rho}(x_{2}) + (1 \leftrightarrow 2)]
$

We can take
$
B = 0
$
and
\be
B^{\mu}(x_{1},x_{2}) = i~\p^{\nu}d_{2}(x_{1} - x_{2})~\p^{\mu}\tilde{u}^{\rho}(x_{1}) \p_{\rho}u_{\nu}(x_{2})
\ee

30) 
$
D = \p_{\mu}\p_{\nu}d_{2}(x_{1} - x_{2})~[ \p_{\rho}u^{\rho}(x_{1}) \p^{\mu}\tilde{u}^{\nu}(x_{2}) + (1 \leftrightarrow 2)]
$

We can take
$
B = 0
$
and
\be
B^{\mu}(x_{1},x_{2}) = i~\p_{\nu}d_{2}(x_{1} - x_{2})~\p^{\mu}\tilde{u}^{\nu}(x_{1}) \p_{\rho}u^{\rho}(x_{2})
\ee

31) 
$
D = \p_{\mu}\p_{\nu}d_{2}(x_{1} - x_{2})~[ \p^{\mu}u^{\nu}(x_{1}) \p_{\rho}\tilde{u}^{\rho}(x_{2}) + (1 \leftrightarrow 2)]
$

We can take
$
B = 0
$
and
\be
B^{\mu}(x_{1},x_{2}) = - i~\p_{\nu}d_{2}(x_{1} - x_{2})~\p^{\mu}u^{\nu}(x_{1}) \p_{\rho}\tilde{u}^{\rho}(x_{2})
\ee

32) 
$
D = \p_{\mu}\p_{\nu}d_{2}(x_{1} - x_{2})~[ \p^{\rho}u^{\mu}(x_{1}) \p_{\rho}\tilde{u}^{\nu}(x_{2}) + (1 \leftrightarrow 2)]
$

We can take
$
B = 0
$
and
\be
B^{\mu}(x_{1},x_{2}) = - i~\p_{\nu}\p_{\rho}d_{2}(x_{1} - x_{2})~[ u^{\nu}(x_{1}) \p^{\mu}\tilde{u}^{\rho}(x_{2})
- \p^{\mu}\tilde{u}^{\nu}(x_{1}) u^{\rho}(x_{2}) ]
\ee

33) 
$
D = \square d_{2}(x_{1} - x_{2})~\p_{\rho}h^{\mu\rho}(x_{1}) \p^{\sigma}h_{\mu\rho}(x_{2})
$

We can take
\be
B(x_{1},x_{2}) = - {i \over 2}~\square d_{2}(x_{1} - x_{2})~[ \p^{\nu}h_{\mu\nu}(x_{1}) \tilde{u}^{\mu}(x_{2})
+ (1 \leftrightarrow 2)]
\ee
and
$
B^{\mu} = 0
$

34) 
$
D = \square d_{2}(x_{1} - x_{2})~[ \p_{\mu}h(x_{1}) \p_{\nu}h^{\mu\nu}(x_{2}) + (1 \leftrightarrow 2)]
$

We can take
\be
B(x_{1},x_{2}) = - i\square d_{2}(x_{1} - x_{2})~[ \p_{\mu}h(x_{1}) \tilde{u}^{\mu}(x_{2}) + (1 \leftrightarrow 2)]
\ee
and
\be
B^{\mu}(x_{1},x_{2}) = i~\p^{\mu}d_{2}(x_{1} - x_{2})~\p_{\nu}\p_{\rho}u^{\rho}(x_{1}) \tilde{u}_{\nu}(x_{2})
- i~d_{2}(x_{1} - x_{2})~\p^{\mu}\p^{\nu}\p^{\rho}u_{\rho}(x_{1}) \tilde{u}_{\nu}(x_{2})
\ee

\newpage
35) 
$
D = \square d_{2}(x_{1} - x_{2})~\p_{\mu}h_{\rho\sigma}(x_{1}) \p^{\mu}h^{\rho\sigma}(x_{2})
$

We can take
$
B = 0
$
and
\be
B^{\mu}(x_{1},x_{2}) = - {i \over 2}~\square d_{2}(x_{1} - x_{2})~[ \p^{\mu}h^{\nu\sigma}(x_{1}) h_{\nu\rho}(x_{2})
+ h_{\nu\rho}(x_{1}) \p^{\mu}h^{\nu\rho}(x_{2}) ]
\ee

36) 
$
D = \square d_{2}(x_{1} - x_{2})~\p_{\mu}h(x_{1}) \p^{\mu}h(x_{2})
$

We can take
$
B = 0
$
and
\be
B^{\mu}(x_{1},x_{2}) = - {i \over 2}~\square d_{2}(x_{1} - x_{2})~[ \p^{\mu}h(x_{1}) h(x_{2}) + h(x_{1}) \p^{\mu}h(x_{2}) ]
\ee

37) 
$
D = \square d_{2}(x_{1} - x_{2})~\p_{\mu}h_{\nu\rho}(x_{1}) \p^{\rho}h^{\mu\nu}(x_{2})
$

We can take
$
B = 0
$
and
\bea
B^{\mu}(x_{1},x_{2}) = - {i \over 2}~[ \p^{\mu}\p^{\nu}d_{2}(x_{1} - x_{2})~\p_{\rho}h_{\nu\sigma}(x_{1}) h^{\rho\sigma}(x_{2}) 
+ \square d_{2}(x_{1} - x_{2})~h_{\nu\rho}(x_{1}) \p^{\rho}h^{\mu\nu}(x_{2}) 
\nonumber\\
- \p_{\nu}d_{2}(x_{1} - x_{2})~\p^{\mu}\p^{\rho}h^{\nu\sigma}(x_{1}) h_{\rho\sigma}(x_{2}) ]
\eea

38) 
$
D = \square d_{2}(x_{1} - x_{2})~[ \p_{\mu}u^{\mu}(x_{1}) \p_{\nu}\tilde{u}^{\nu}(x_{2}) + (1 \leftrightarrow 2)]
$

We can take
$
B = 0
$
and
\be
B^{\mu}(x_{1},x_{2}) = - i~\p^{\mu}d_{2}(x_{1} - x_{2})~\p_{\nu}u^{\nu}(x_{1}) \p_{\rho}\tilde{u}^{\rho}(x_{2})
+ i~d_{2}(x_{1} - x_{2})~\p^{\mu}\p^{\nu}u_{\nu}(x_{1}) \p_{\rho}\tilde{u}^{\rho}(x_{2})
\ee

We have proved the formula from the statement. If we make 
$
d_{2} \mapsto d_{2}^{F}
$
in the expressions
$
B
$
and
$
B^{\mu}
$
we obtain the expressions
$
B^{F}
$
and
$
B^{F,\mu}
$
so the chronological product
$
T_{(1)}(T(x_{1}),T(x_{2}))
$
is also a relative coboundary.
$\qed$

\begin{rem}
We emphasize that in the preceding proof we did not need the explicit expressions for the coefficients of the $38$ monomials
from (\ref{D-one-loop-a}). Moreover, it can be proved that a generic expression of the type (\ref{D-one-loop-a}) has $126$
generic terms and all of them are coboundaries, as above.
\end{rem}

Let us try to establish cohomological formulas as in the preceding theorem for the loop contributions of the other chronological products.
We easily get from (\ref{gau1}) that 
\be
d_{Q}D_{(1)}(T(x_{1}),T(x_{2}))
= i~[ \p_{\mu}^{1}D_{(1)}(T^{\mu}(x_{1}),T(x_{2})) - (1 \leftrightarrow 2)]
\ee
If we use the formula obtained in the preceding theorem we obtain:
\be
\p_{\mu}^{1}~[ D_{(1)}(T^{\mu}(x_{1}),T(x_{2})) - d_{Q}B^{\mu}(x_{1},x_{2})]  = (1 \leftrightarrow 2).
\label{cocycle}
\ee

We cannot apply Poincar\'e lemma because the cochains space has the particular form (\ref{cochain})
and the usual homotopy formula used to prove Poincar\'e lemma does not preserve this form. In the absence of a better idea we can 
proceed by a brute force analysis. We consider cochains 
$
D^{\mu}(x_{1},x_{2})
$
of the form (\ref{cochain}) of ghost number $1$, canonical dimension $6$, Lorentz covariant and verifying the cocycle condition
\be
\p_{\mu}^{1}~D^{\mu}(x_{1},x_{2}) = (1 \leftrightarrow 2);
\label{cocycle-1}
\ee
we try to see if such an expression is a coboundary i.e. of the fom
\be
D^{\mu}(x_{1},x_{2}) = \p_{\nu}^{1}B^{[\mu\nu]\emptyset}(x_{1},x_{2}) - \p_{\nu}^{2}B^{[\mu][\nu]}(x_{1},x_{2})
\ee
with the expressions
$
B^{[\mu\nu]\emptyset}
$
and
$
B^{[\mu][\nu]}
$
also cochains of the form (\ref{cochain}). A hard computation shows that the cohomology is non-trivial. Every cocycle is cohomologous 
with a cochain of the form 
\be
D_{0}^{\mu}(x_{1},x_{2}) = [ \p^{\mu}u^{\nu}(x_{1})~\p_{\nu}\p_{\rho}\p_{\sigma}h^{\rho\sigma}(x_{2})
- u_{\nu}(x_{1})~\p^{\mu}\p^{\nu}\p^{\rho}\p^{\sigma}h^{\rho\sigma}(x_{2}) ] - ( 1 \leftrightarrow 2)
\ee
which is a non-trivial cocycle. So, without explicit computations, we can say that
\be
D_{(1)}(T^{\mu}(x_{1}),T(x_{2})) = d_{Q}B^{\mu}(x_{1},x_{2}) 
+ i~[\p_{\nu}^{1}B^{[\mu\nu]\emptyset}(x_{1},x_{2}) - \p_{\nu}^{2}B^{[\mu][\nu]}(x_{1},x_{2})] + \lambda~D_{0}^{\mu}(x_{1},x_{2})
\ee

If we want to prove that 
$
\lambda = 0
$
such that
$
D_{(1))}(T^{\mu}(x_{1}),T(x_{2}))
$
is a relative coboundary, then one must use the explicit expressions. Indeed, one can prove that neither 
$
D_{(1))}(T^{\mu}(x_{1}),T(x_{2}))
$
or 
$
d_{Q}B^{\mu}(x_{1},x_{2})
$
appearing in (\ref{cocycle}) have contributions proportional to
$
D_{0}^{\mu}(x_{1},x_{2}). 
$

We iterate the preceding argument in the sector of ghost $2$ and we obtain in the same way a formula of the type (\ref{cocycle-1}) 
i.e. a cocycle condition of the type

\be
\p_{\nu}^{1}~D^{[\mu\nu]\emptyset}(x_{1},x_{2}) - \p_{\nu}^{2}D^{[\mu][\nu]}(x_{1},x_{2})  = 0
\ee
where the expressions $D$ are cochains of the form (\ref{cochain}). We want to establish
if these expressions are coboundaries i.e. of the form:
\bea
D^{[\mu\nu],\emptyset}(x_{1},x_{2}) = \p_{\rho}^{1}B^{[\mu\rho][\nu]}(x_{1},x_{2}) + \p_{\rho}^{2}B^{[\mu\nu][\rho]}(x_{1},x_{2}) 
\nonumber\\
D^{[\mu],[\nu]}(x_{1},x_{2}) = \p_{\rho}^{1}B^{[\mu\rho][\nu]}(x_{1},x_{2}) + (x_{1} \leftrightarrow x_{2}, \mu \leftrightarrow \nu)
\eea
with $B$ cochains of the form (\ref{cochain}) and of ghost number $2$. By a long computation one can establish that this is true.
As a consequence, the expression
$
D^{IJ}_{(1)}(x_{1},x_{2}),~|I| + |J| = 2
$
is a relative coboundary.

Finally, because the expression
$
D_{(2)}(x_{1},x_{2})
$
is proportional to
$
\square^{2}d_{3}
$
it is easy to put it in the form of a coboundary of the form
$
\delta B.
$
\newpage
\section{Conclusions}
We have proved that the loop contributions to the causal commutator
$
D^{IJ}_{(1)}
$
are of the form
$
\bar{s}B
$
in the pure gravity case. Because the expressions $B$ have also causal support
this property stays true after causal splitting.
This means that, in the second order of the perturbation theory, the physical contributions is the tree contribution
which correspond to the classical theory. We conjecture that this result stays true in all orders of the perturbation
theory. This conjecture might be true in the non-perturbative case also.

We remark that our approach differs from some other recent approaches to quantum gravity without ghosts. In these aproaches
one cannot establish a super-renormalizability result as above, so at best, one can treat quantum gravity as an effective theory.

\end{document}